\documentclass[%
 reprint,
 amsmath,amssymb,
 aps,nofootinbib,   
]{revtex4-1}

\usepackage{bm}
\PassOptionsToPackage{linktocpage}{hyperref}
\usepackage[hyperindex,breaklinks]{hyperref}
\usepackage{enumitem}
\usepackage{slashed}

\usepackage{xcolor}
\usepackage{float}

\usepackage{array}
\usepackage{mathtools}

\usepackage{etoolbox}

\begin{document} 

\title{Cosmology from quantum information}
\author{C\'esar G\'omez} 
\affiliation{Instituto de F\'{i}sica Te\'orica UAM-CSIC, Universidad Aut\'onoma de Madrid, Cantoblanco, 28049 Madrid, Spain.}

\author{Raul Jimenez} 
\affiliation{ICC, University of Barcelona, Marti i Franques 1, 08028 Barcelona, Spain.}
\affiliation{ICREA, Pg. Lluis Companys 23, Barcelona, E-08010, Spain.}


\begin{abstract}
We describe inflation in terms of a time dependent quantum density matrix with time playing the role of a stochastic variable. Using a quasi-de Sitter model we compute the corresponding quantum Fisher information function as the second derivative of the relative entanglement entropy for the density matrix at two different times. Employing standard quantum estimation theory we evaluate the minimal variance of quantum scalar fluctuations that reproduces the power spectrum and the corresponding tilt in the slow roll limit. The Jeffreys prior associated with such Fisher information can be used to define the probabilities on the set of initial conditions defined by the slow roll parameter $\epsilon$ and the initial Shannon information.
\end{abstract}

\maketitle
The current cosmological model is the $\Lambda CDM$, whose seven free parameters are well constrained, at the \% level, by the data from the Planck mission \cite{Planck18}. This model provides a remarkable description of a wealth of cosmological observations. This model in its early phase, postulates an epoch of accelerated expansion that leads to a very large and homogeneous Universe. After this super-cooled expansion phase, reheating takes place; in this phase of reheating the inflaton converts its energy into particles via resonant coupling to the standard model of particle physics. This hot big-bang phase further leads to a matter dominated Universe. This early inflationary epoch is crucial to explain the flatness and the present homogeneity of the observed universe \cite{Starobinsky,Guth,Linde}.  The quantum origin of the large scale structure is explained on the basis of the classicalization of small quantum fluctuations that leave the horizon during the inflationary epoch and re-enter during the hot big-bang period \cite{Mukhanov,Hawking,Albrecht,GuthPi}. This is achieved if the time evolution of the horizon changes from shrinking during the inflationary epoch to increasing after reheating. While we have not yet measured the smoking gun of an exponential phase of expansion in the early Universe (the much expected primordial B-mode polarization of the cosmic microwave background), there are two remarkable measurements that strongly support inflation as the correct picture to describe the early Universe: the existence of super-horizon fluctuations~\cite{WMAP03} and the fact that the power spectrum of matter fluctuations is nearly scale-invariant and has a red tilt~\cite{Planck18}.

There is however not a definite picture of the physics behind inflation. The usual approach is to postulate a potential from a hypothetical scalar field and adjust this potential in order to fit the observational constraints.  The recent Planck~\cite{Planck18} analysis already constrain this potential to be very flat. In this letter we try a different approach to describe the epoch of inflation and the early Universe. 

Our guiding principle will be quantum information and more precisely quantum estimation theory (see Ref.~\cite{Paris}) \footnote{For recent applications of quantum estimation theory in the same spirit see Ref.~\cite{Gomez1,Gomez2,Bhattacharyya:2020rpy} .}. From the point of view of a quantum mechanical description, we are interested in the time dependence of a set of cosmological observables. Since time itself in quantum mechanics is not an observable, we interpret it as a stochastic variable and we encode the information about time dependence in terms of an Fisher function. Through its relation with the relative entanglement entropy (for a review see e.g. \cite{RE})  that defines a natural distance between the quantum states of the early universe at different times, we introduce a simple quasi de Sitter model to estimate the quantum Fisher function.

Using the standard relations between the Fisher function and the variance of the associated stochastic quantity, we reproduce some of the most basic results about inflation; in particular, bounds on the duration of inflation as well as the power spectrum of scalar quantum fluctuations. This is done without assuming any particular form for the inflaton potential. An interesting and potentially deep output of this approach, is the important role of the saturation of minimal variance on the basis of the quantum Cramer-Rao inequality that sets the variance of cosmological fluctuations as given by the power spectrum at the point of horizon exit.

As a first approximation we can try to describe the inflationary epoch in terms of a time dependent Hubble parameter $H$ that is postulated to decrease i.e. $\dot H\leq 0$. Quantum mechanically we can describe the state of the universe at time $t$ using a quantum density matrix $\rho(t)$ that we can define by taking the trace of the quantum state over the whole region outside the horizon at time $t$ \footnote{Assuming the quantum state of the universe at a given time is $|\Psi\rangle$ we define this density matrix formally as $Tr |\Psi\rangle \langle \Psi|$ with the trace over the region outside the horizon at that time. In what follows we will not need a concrete characterization of the state $|\Psi\rangle$}. Now we can define the relative entanglement entropy between two different times as
\begin{equation}\label{def}
S(t;t_0) \equiv Tr(\rho(t) \ln \frac{\rho(t)}{\rho(t_0)})
\end{equation}
for $t_0$ the initial time. As it is well known in information theory $S(t;t_0)$ is positive definite and leads to the first law of entanglement, namely
\begin{equation}
\dot S(t) =0
\end{equation}
If we Taylor expand around $t_0$ we get
\begin{equation}
S(t_0+\delta_t;t_0) = S(0) + \frac{1}{2}\delta_t^2 F(t_0)
\label{eq:entropy}
\end{equation}
with $F(t_0)$ the quantum Fisher information at the initial time $t_0$. Note from the definition (\ref{def}) that $S(0)=0$. We shall characterize the initial condition of our cosmology in terms of the value of the initial Fisher information. In statistical terms, the time $t$ is working like a stochastic variable i.e. time is just defining the variable space and the Fisher function defines the metric on this space. 

The relative entropy can be represented as
\begin{equation}
S(t;t_0)= Tr(\rho(t) {\cal{H}}) - Tr(\rho(t_0){\cal{H}}) - S(\rho(t)) + S(\rho(t_0))
\end{equation}
for $S(\rho)$ the standard von Neumann entropy and ${\cal{H}}$ the modular or entanglement Hamiltonian. If the density matrix is hermitian and positive semidefinite it can be always expressed as $\rho = \frac{e^{-{\cal{H}}}}{tr(e^{-{\cal{H}}})}$ for ${\cal{H}}$ some hermitian operator. This is the operator defining the modular Hamiltonian. In essence the modular Hamiltonian represents the density matrix as a formal canonical matrix. In general to find the modular Hamiltonian for a generic density matrix is a very difficult task. In our case we will profit from representing the quasi de Sitter density matrix as thermal for the Gibbons-Hawking temperature~\cite{Gibbons}. Once we assume $t$ close to $t_0$ and we choose for $\rho(t_0)$ a canonical density matrix at temperature $T$, the relative entropy can be represented as
\begin{equation}\label{free}
 S(t;t_0) = \frac{1}{T}(F(\rho(t))-F(\rho(t_0))) 
 \end{equation}
 with $F(\rho)$ the free energy i.e. $(E-sT)$ for $E$ the energy and $s$ the entropy. We will use this representation in order to estimate the relative entropy. Moreover the quantum Fisher function can be written as
\begin{equation}
F(t) = Tr(\rho(t) L_t^2)
\label{eq:Fisher}
\end{equation}
for $L_t$ the Lyapunov operator driving the time evolution of $\rho$ \footnote{This operator simply represents the symmetric logarithmic derivative and it is defined by the Lyapunov equation $\frac{d\rho}{dt} = \frac{L_t \rho +\rho L_t}{2}$.}. Models of inflation are in fact ways to model $L_t$ and consequently also ${\cal{H}}$. In what follows we will use to estimate the Fisher function a simple model where we only assume that $H$ depends on time and that $\dot H$ is negative. 

Before going into the detailed computation, let us highlight the key point of our approach in more qualitative terms. In general relativity, canonical quantizations lead to the Wheeler-de Witt equation which does not involve time. To recover standard time evolution it requires to introduce a clock field, let us say $\phi$ \footnote{See for instance the discussion in Ref.~\cite{Nima}.}. In cosmology this clock field is just the inflaton. Once we have defined, using this clock, time evolution, we can formally compute the associated Lyapunov operator that depends on the corresponding clock. This allows us to compute the associated quantum Fisher function $F_{\phi}$ using (\ref{eq:Fisher}). Our main claim is that the corresponding power spectrum $\Delta_{\phi}^2$ is just determined by the inverse of the quantum Fisher function
\begin{equation}
\Delta_{\delta \phi}^2 \sim \frac{1}{F_{\phi}}
\end{equation}
In other words, once we introduce a clock to parametrize the time dependence, the corresponding quantum variance defines the power spectrum of fluctuations of the field used to define the clock.

In what follows we will estimate the cosmological quantum Fisher function using its relation with the relative entropy (\ref{eq:entropy}). We shall define $S(t;t_0)$ using (\ref{free}) as $(E(t)/T - s(t))- (E(t_0)/T - s(t_0))$ for $s(t), E(t), T$ respectively entropy, energy and effective temperature. Moreover we will take the same $T$ as defined by the canonical density matrix at time $t$ \footnote{Physically  this means that locally we assume that the system tends to equilibrium with time. More precisely we assume that the thermal equilibrium matrix is the one at $t$ for $t$ in the past of $t_0$.}
\begin{equation}
\rho(t) = \frac{e^{E(t)/T}}{Tr (e^{E(t)/T})}
\end{equation}

 For a given time characterized by $H(t)$ we define, in Planck units,
\begin{equation}\label{relative}
S(t;t_0) =\frac{(a-b)}{H^2} - \frac{a}{HH_0} + \frac{b}{H_0^2}
\end{equation}
for $a$, $b$ two parameters with units of $M_P^2$ and with $H_0$ the Hubble rate at the initial time. The logic of this model of the relative entropy is based on assuming the following quasi de Sitter relations: $T\sim H$, $E(t) \sim \frac{H(t)^2}{H(t)^3}$ and $s(t) \sim \frac{1}{H(t)^2}$ all of them in natural Planck units. The temperature $T$ is simply defined as the Gibbons-Hawking temperature~\cite{Gibbons} for a de Sitter with Hubble parameter $H$. Infinitesimally around a reference point we define the difference in free energy keeping the same $T$ and changing the entropy and the energy accordingly with the de Sitter relations. This approximation leads to (\ref{relative}) as a phenomenological model of the relative entropy. 

Note that~\ref{relative} satisfies $S(t_0;t_0)=0$ and that this model of relative entropy is well defined at the inflationary epoch since $S(t;t_0)$ is always positive for any time $t>t_0$. Secondly, in order to satisfy the first law of entanglement, namely $\dot S=0$, in~\ref{relative}, $a=2b$. Now we can easily compute the Fisher function:
\begin{equation}
F(t_0) = 2b \epsilon^2
\end{equation}
where we have introduced the slow-roll parameter
\begin{equation}
\epsilon = -\frac{\dot H}{H^2}.
\end{equation}
We have only used the assumption that $H$ changes with time and we define $\epsilon$ in a way that is not assuming any underlying inflaton potential model. Thus we observe that the slow-roll simply generates a non vanishing Fisher function. Using now the Cramer-Rao inequality, this finite Fisher function can be used to define the uncertainty on time, which is bounded by the minimal uncertainty
\begin{equation}\label{one}
\Delta_{\delta t} \sim \frac{1}{\sqrt{2b}\epsilon}
\end{equation}
As it is well known this relation is the statistical version of the quantum energy time uncertainty principle. Saturating the Cramer-Rao inequality is equivalent to minimizing the corresponding uncertainty.

What we can call the minimal statistical duration of inflation is determined by the initial value $\epsilon$ by (\ref{one}). This is measured in Planck units. To change to Hubble units, and taking $2b$ order one, we get
\begin{equation}
\Delta_{\delta t} = \frac {{\cal{N}}}{\sqrt{N_0}}
\end{equation}
Hubble times. Here ${\cal{N}}$
is the number of e-foldings and $N_0\equiv \frac{M_P^2}{H_0^2}$. So if we tune the initial Hubble to be Planckian, the saturation of Cramer-Rao sets the duration of inflation. In other words the physical duration of inflation, given in Hubble units as the number of e-foldings, let us say $\Delta_{phys}$, is given by
\begin{equation}
\Delta_{phys} \sim \sqrt{N_0} \Delta_{\delta t}
\end{equation}

In summary, from the information point of view, the minimal step in the cosmological evolution \footnote{In computational language the time of the elementary "cosmological gate".} is set by the initial Fisher information and is given by $\Delta_{\delta t}$. The number of cosmological steps that we can identify as the complexity of the whole inflation process is determined by $\sqrt{N_0}$ and therefore is stablished by the initial Shannon information.

 Note also that $\epsilon > 1$, which is the standard way to set the end of inflation, makes no sense, from the information theory point of view, since it leads to time uncertainties that are sub-Planckian i.e. smaller than the Planck time \footnote{In essence we assume that errors in estimating the statistical parameter defining time cannot be smaller than a Planck unit.}. This is equivalent to say that one cannot model inflation with $\epsilon >1$ by using only information theory arguments.

The information content of the initial state of the universe is, therefore, a Shannon information $N_0$ and a Fisher information. If we assume an inflationary epoch with decreasing Hubble parameter, the Fisher information sets the corresponding rate of change. The limiting case $\epsilon =0$ leads to vanishing Fisher information. Using the qualitative relation between Fisher and Shannon for the Gaussian case \cite{Stam}  and assuming it can be generalized, will imply that $\epsilon=0$ is only consistent with $N_0=\infty$.

Let us now briefly discuss the crucial issue of generation of seeds for galaxy formation as  quantum fluctuations.  Until now we have discussed the Fisher function associated to time as a stochastic variable and discover that $F$ goes like $\epsilon^2$. This means that the corresponding variance goes like 
\begin{equation}
\Delta_{\delta t}^2 \sim \frac{1}{\epsilon^2}
\end{equation}
Physically one can figure out the meaning of this expression as defining the variance of time on a surface of constant energy density.  

This effective variance in time induces in general relativity the curvature fluctuations. In essence, the quantum fluctuations in inflation have as origin that inflation ends at different times in different regions of space. To describe this phenomena we simply use the variance of time (interpreted as a stochastic variable) induced by the initial Fisher function.

 If we are interested in the expression of these fluctuations for an hypothetical inflaton field $\phi$ we can use the relation $\delta \phi = \dot \phi \delta t$. The quantity that is now dependent on the particular model of inflation is $\dot \phi$.  In the slow roll approximation and in Planck units we can use
\begin{equation}
\dot \phi^2 \sim H^2 \epsilon.
\end{equation}
This leads to the variance of $\delta \phi$ corresponding to the power spectrum

\begin{equation}
\Delta_{\delta \phi}^2 \sim \dot \phi^2 \Delta_{\delta t}^2 \sim \frac{H^2}{\epsilon}
\end{equation}

which is the expected result for the power spectrum of scalar perturbations~\cite{Mukhanov}. In other words, the Fisher function for $\phi$, i.e., when we take the inflaton scalar field as a stochastic variable, controls the power spectrum of curvature perturbations through the Cramer-Rao bound
\begin{equation}
\Delta_{\delta \phi}^2 \sim \frac{1}{F_{\phi}}.
\end{equation}
with $F_{\phi}= \frac{\epsilon}{H^2}$. 

At this point it could be worth to make a comment on the relation between this statistical derivation and the standard derivation of the power spectrum for the curvature perturbations. In this second case the time fluctuations at the end of inflation, on a field hypersurface, let us say $\delta t$, induce curvature perturbations of order $H\delta t$ and therefore a power spectrum that goes like $H^2 (\delta t)^2$ with $\delta t \sim \frac{1}{\sqrt{\epsilon}}$. We are instead representing the curvature power spectrum as $\dot \phi^2 (\Delta t )^2$ with $\Delta t$ defined by the Fisher function as the total duration of the process of inflation i.e. $\frac{1}{\epsilon}$.  Thus the power spectrum of curvature perturbations is simply defined as $(\Delta \phi)^2 = \dot \phi ^2 (\Delta t )^2$ or equivalently $(\Delta \phi)^2 \sim \langle \delta \phi \delta \phi \rangle \Delta(t)$ with the standard value of $\langle \delta \phi \delta \phi \rangle \sim H^2$.

It is interesting to observe that the saturation of the Cramer-Rao bound leads to  the power spectrum at the scale of horizon exit in agreement with our former discussion on the duration of inflation. More precisely, it saturates Cramer-Rao for $\Delta \phi = \frac{H}{\sqrt {\epsilon}}$ that corresponds to 
$\Delta t = \frac{1}{\epsilon}$. In essence, the modes freeze when they saturate the minimal value for the variance.  In order to estimate the tilt 
\begin{equation}
\frac{d\ln(\Delta^2)}{d\ln(k)}
\end{equation}
for $k$ the value at the horizon exit we use that the horizon exit takes place after a time order $\frac{1}{\epsilon}$ in Hubble units. Using now that $\Delta^2 \sim \frac{1}{\epsilon}$ we get in this approximation
 \begin{equation}
\frac{d\ln(\Delta^2)}{d\ln(k)} = n_s-1 \sim - \epsilon
\end{equation}
in qualitative agreement with the expected result~\cite{Mukhanov,Planck18}.

Finally, we can use the former Fisher function $F= \frac{\epsilon}{H^2}$
to define a Jeffreys prior probability on cosmological initial conditions 
\begin{equation}
J = \sqrt{F}= \sqrt{\epsilon} \sqrt{N_0}
\end{equation}
 From this point of view the prior probability of no inflation i.e. of $\epsilon=0$ is just zero and the highest probability appears when  the number of e-foldings $\frac{1}{\epsilon}$ is of the order of the initial Shannon information. We leave for a future work a more careful analysis of this qualitative statistical argument and its relation to Bayesian statistics.
\\

\acknowledgments
We thank Alan Heavens and Shao-Jiang Wang for useful comments. The work of CG was supported by grants SEV-2016-0597, FPA2015-65480-P and PGC2018-095976-B-C21. The work of RJ is supported by grant PGC2018-098866-B-I00.

\end{document}